\RequirePackage{fix-cm}
\documentclass[twocolumn]{svjour3}          
\smartqed  
\usepackage{graphicx}
\usepackage{amsmath}
\usepackage{amssymb}
\usepackage{upgreek}
\usepackage[hyphens]{url}
\usepackage{tabularx}
\usepackage{bigstrut}
\usepackage{textcomp}
\usepackage{gensymb}
\usepackage[english]{babel}
\usepackage{multirow}
\usepackage{rotating}
\newcommand\tabrotate[1]{\begin{turn}{90}\rlap{#1}\end{turn}}

\usepackage{varwidth}
\newcommand\tabvarwidth[2][3cm]{\begin{varwidth}[b]{#1}\centering #2\end{varwidth}}

\usepackage{array}
\newcolumntype{L}[1]{>{\raggedright\let\newline\\\arraybackslash\hspace{0pt}}p{#1}}
\newcolumntype{C}[1]{>{\centering\let\newline\\\arraybackslash\hspace{0pt}}p{#1}}
\newcolumntype{R}[1]{>{\raggedleft\let\newline\\\arraybackslash\hspace{0pt}}p{#1}}

\journalname{arXiv submission}
\begin{document}

\title{A concept for Lithography-free patterning of silicon heterojunction back-contacted solar cells by laser processing}
\author{B. Turan \and K. Ding \and S. Haas}
\institute{B. Turan \at
              Forschungszentrum J\"ulich \\
              IEK5 -- Photovoltaik\\
              Tel.: +49-2461-61-9089\\
              Fax: +49-2461-61-3735\\
              \email{b.turan@fz-juelich.de}}           

\date{Uploaded: 09.06.2015}
\maketitle

\begin{abstract}
Silicon heterojunction (SHJ) solar cells with an interdigitated back-contact (IBC) exhibit high conversion efficiencies of up to 25.6\%. However, due to the sophisticated back-side pattern of the doped layers and electrode structure many processing and patterning steps are required. A simplification of the patterning steps could ideally increase the yield and/or lower the production costs. We propose a patterning approach for IBC SHJ solar cells free of any photo-lithography with the help of laser-induced forward transfer (LIFT) of the individual layer stacks to create the required back-contact pattern. The concept has the potential to lower the number of processing steps significantly while at the same time giving a large degree of freedom in the processing conditions optimization of emitter and BSF since deposition of the intrinsic/doped layers and processing of the wafer are all independent from each other.     


\end{abstract}

\noindent
\section{Introduction}
\label{intro}
\begin{figure}
	\centering
	\includegraphics[width=0.99\columnwidth]{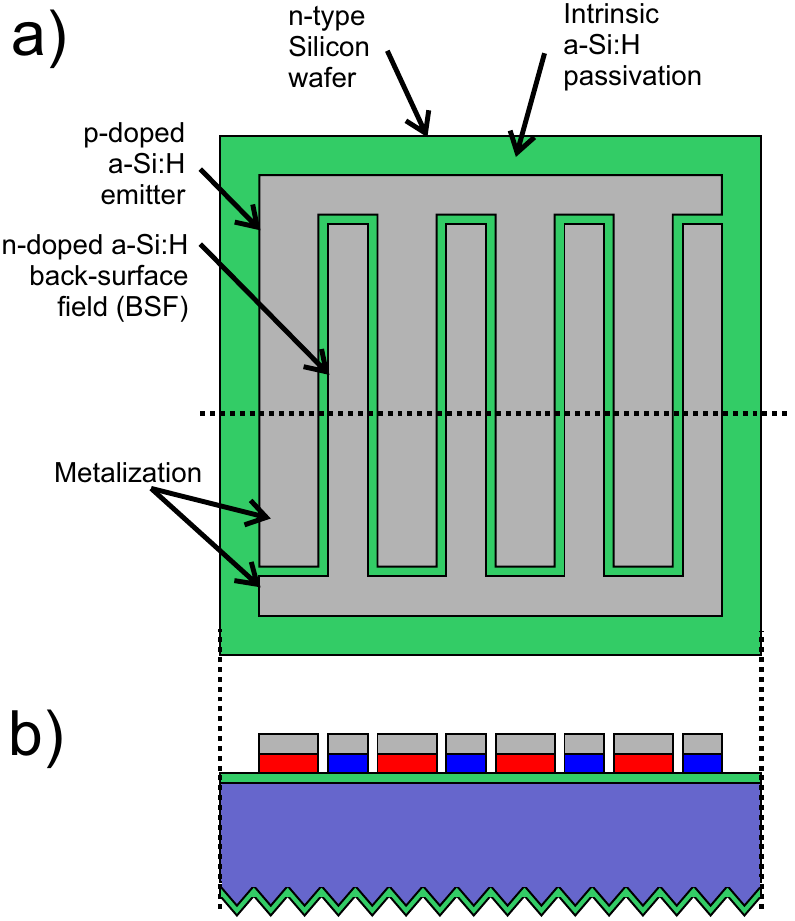}
	\caption{Sketch of the back-side structure of a silicon heterojunction solar cell (SHJ) with an interdigitated back-contact (IBC). Top-view in a) and side-view through dashed line in b). The emitter (in blue) and back-surface field (shown in red) region are formed in an interdigitated spatial structure on top of a passivated silicon wafer with a small gap in-between both regions. On top of both doped amorphous silicon regions a metalization is deposited for the extraction of the generated charge carriers.}
	\label{fig:sketch}
\end{figure}

Interdigitated back-contacted silicon heterojunction solar cells exhibit the highest conversion efficiencies for silicon based solar cells to date with values up to 25.6\% \cite{Masuko2014}. This solar cell topology combines the high output voltages of silicon heterojunction solar cells (SHJ) (often also called heterojunction with intrinsic thin layer (HIT))~\cite{Hamakawa1983}\cite{Wolf2012} thanks to their efficient passivation of the crystalline silicon surface, with the high output currents of interdigitated back contacted solar cells (IBC)~\cite{Lammert1977}, which eliminates optical shading by grid electrodes on the front side of the solar cell~\cite{Serreze1978}. Yet, back-contacted silicon solar cells require a structure layout on the back side where rear emitter and the back surface field (BSF) are defined spatially in, for example, an interdigitated manner. Figure~\ref{fig:sketch} shows schematically the back-side structure of a IBC SHJ solar cell.

Various electro-optical requirements and constraints determine the optimal geometrical fill-factors of emitter and BSF. Furthermore, the gap distance in-between both regions should be as small as possible~\cite{Reichel2011} without a short-circuit by the subsequent metalization between emitter and BSF.

For the generation of the back-side structure different patterning techniques are reported in literature. These are most prominently photo-lithography, ink-jet resist printing~\cite{Stuwe2015}, shadow masking or laser doping in the case of IBC cells with a homo junction. All these techniques have different advantages and disadvantages concerning the process stability, costs, complexity, and spatial resolution. Table~\ref{tab:comp} shows a basic qualitative comparison of some of the reported approaches.

\begin{table}[htbp]
	\centering
	\caption{Comparison of different technologies for the patterning and creation of the back-contact structure of IBC SHJ solar cells. Ranking is done with four marks. Two plus signs is synonymous with a large advantage versus the other methods while two minus signs correspond to a low evaluation of this approach in comparison.}
	\begin{tabular}{p{0.06\columnwidth}|p{0.21\columnwidth}|p{0.16\columnwidth}|p{0.18\columnwidth}|p{0.15\columnwidth}}
		\hline
		\multicolumn{5}{c}{\textbf{Technology}} \\
		\hline
					& Photo-\newline lithography & Inkjet & Shadow-\newline masking & LIFT/\newline LDW \\
		\hline
			\multirow{2}{*}{\rule{0pt}{55pt}\tabrotate{\tabvarwidth{\textbf{Equipment\newline costs}}}}
			&\begin{minipage}[c]{0.5cm}
				\includegraphics[width=0.5cm]{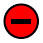}\includegraphics[width=0.5cm]{Minus.pdf}
			\end{minipage}
			& \begin{minipage}[c]{0.5cm}
				\includegraphics[width=0.5cm]{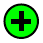}
			\end{minipage}
			& \begin{minipage}[c]{0.5cm}
				\includegraphics[width=0.5cm]{Plus.pdf}\includegraphics[width=0.5cm]{Plus.pdf}
			\end{minipage}
			&\begin{minipage}[c]{0.5cm}
				\includegraphics[width=0.5cm]{Minus.pdf}
			\end{minipage} \\
		    
		    & (Contact-) Mask aligner\vspace{15pt} & Printing tool \vspace{15pt}& Masking PECVD \vspace{15pt}& Laser scribing tool \vspace{15pt}\\
		\hline
		 	\multirow{2}{*}{\rule{0pt}{50pt}\tabrotate{\tabvarwidth{\textbf{Process\newline costs}}}}
		 	& \begin{minipage}[c]{0.5cm}
		 		\includegraphics[width=0.5cm]{Minus.pdf}\includegraphics[width=0.5cm]{Minus.pdf}
		 	\end{minipage}
		 	& \begin{minipage}[c]{0.5cm}
		 		\includegraphics[width=0.5cm]{Minus.pdf}
		 	\end{minipage}
		 	& \begin{minipage}[c]{0.5cm}
		 		\includegraphics[width=0.5cm]{Plus.pdf}\includegraphics[width=0.5cm]{Plus.pdf}
		 	\end{minipage}
		 	&\begin{minipage}[c]{0.5cm}
		 		\includegraphics[width=0.5cm]{Plus.pdf}
		 	\end{minipage} \\
		 			
		 	& Wet-chemical, resist strip, and lift-off & Wet-chemical, lift-off & Mask renewal, cleaning & Donor substrate preparation \\
		\hline
		 	\multirow{2}{*}{\rule{0pt}{55pt}\tabrotate{\tabvarwidth{\textbf{Typical feature-size}}}}
		 	& \begin{minipage}[c]{0.5cm}
		 		\includegraphics[width=0.5cm]{Plus.pdf}\includegraphics[width=0.5cm]{Plus.pdf}
		 	\end{minipage}
		 	& \begin{minipage}[c]{0.5cm}
		 		\includegraphics[width=0.5cm]{Minus.pdf}
		 	\end{minipage}
		 	& \begin{minipage}[c]{0.5cm}
		 		\includegraphics[width=0.5cm]{Minus.pdf}\includegraphics[width=0.5cm]{Minus.pdf}
		 	\end{minipage}
		 	&\begin{minipage}[c]{0.5cm}
		 		\includegraphics[width=0.5cm]{Plus.pdf}
		 	\end{minipage} \\
		 	
		 	& $F\leq$1\,$\micro$m \vspace{30pt}& $F\leq$50\,$\micro$m \vspace{20pt}& $F\leq$100\,$\micro$m \vspace{20pt}& $F\leq$10\,$\micro$m \vspace{20pt}\\	
		 	
		 \hline
		 	\multirow{2}{*}{\rule{0pt}{85pt}\tabrotate{\tabvarwidth{\textbf{Mass-production/\newline Scalability}}}}
		 	& \begin{minipage}[c]{0.5cm}
		 		\includegraphics[width=0.5cm]{Minus.pdf}
		 	\end{minipage}
		 	& \begin{minipage}[c]{0.5cm}
		 		\includegraphics[width=0.5cm]{Plus.pdf}
		 	\end{minipage}
		 	& \begin{minipage}[c]{0.5cm}
		 		\includegraphics[width=0.5cm]{Minus.pdf}\includegraphics[width=0.5cm]{Minus.pdf}
		 	\end{minipage}
		 	&\begin{minipage}[c]{0.5cm}
		 		\includegraphics[width=0.5cm]{Plus.pdf}\includegraphics[width=0.5cm]{Plus.pdf}
		 	\end{minipage} \\
		 			
		 	& Parallel patterning, mask overlay alignment\vspace{10pt}& Serial patterning, print tool alignment\vspace{10pt}& Parallel patterning, mask overlay alignment during PECVD \vspace{10pt}& Serial process, no alignment required \vspace{10pt}\\		
	\end{tabular}%
	\label{tab:comp}%
\end{table}%

In every case multiple processing steps are required which need to be aligned well to each other to ensure a small gap width between the emitter and the BSF.

This paper introduces a concept for a lithography- and contact-free two-step patterning approach for IBC SHJ solar cells. A well-known patterning technique from literature called laser-induced forward transfer (LIFT) or laser direct writing (LDW) is used to transfer the whole functional layer-stack in one step from a donor onto the receiver.  

An example process chain will be introduced, different processing setups will be compared concerning their suitability, and known problems from other LIFT applications will be evaluated with regards to the proposed concept for IBC SHJ solar cells.

\section{Concept}
\begin{figure*}
\centering
\includegraphics[width=0.99\linewidth]{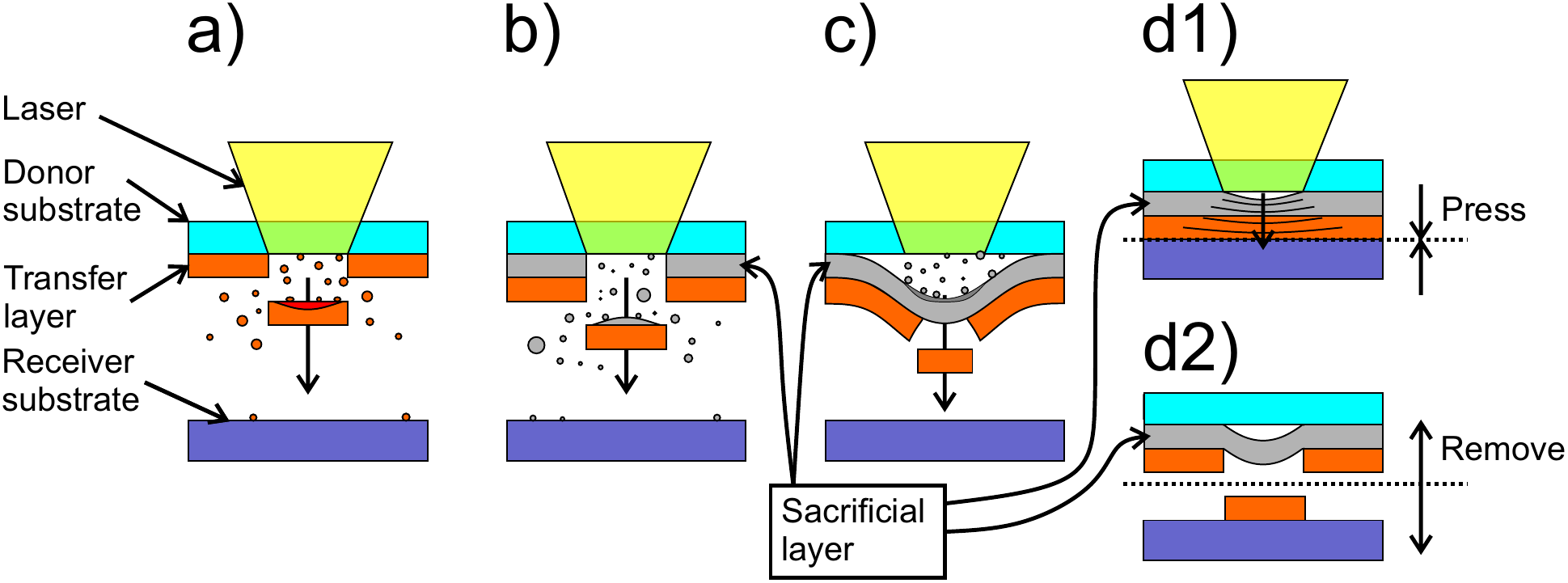}
\caption{Sketch of different laser-induced forward transfer concepts (LIFT). Depicted setups from left to right: a) Simple approach with evaporation of the transfer layer at the donor substrate/transfer layer interface. b) Evaporation of a special sacrificial layer to build the kinetic energy needed to accelerate the transfer layer. c) Same topology as shown in b) but with merely bulging and blistering of the sacrificial layer by incomplete evaporation. d1) and d2) Two step process with donor and receiving substrate in close proximity (contact pressure). After illumination by the laser beam, both substrates are removed from another and the transfer layer remains on the receiver. The small circles illustrate debris that can form during the processing.}
\label{fig:lift_sketch}
\end{figure*}

A schematic sketch of possible LIFT or LDW processes are shown in Figure \ref{fig:lift_sketch}. There are many more approaches for a vast spectrum of applications from MEMS devices~\cite{Birnbaum2010}, pixel transfer for OLEDs~\cite{Feinaeugle2012},~\cite{Cho2012}, gentle transfer of living cells~\cite{Kattamis2007}, to printing of liquid solutions~\cite{Boutopoulos2013}. Good overviews on the topic can be found in~\cite{Arnold2007}, and ~\cite{Nagel2012}. However, here only the methods that are suitable for the specific task of LIFT for IBC SHJ will be discussed.

In every approach shown in Figure~\ref{fig:lift_sketch} two main aspects are found. There is always a donor substrate and a receiver substrate required. Laser processing is used to transfer material from the donor to the receiver. Furthermore, since the process is induced by the laser in each approach the donor substrate is transparent to the laser light, meaning that no absorption of the laser pulse energy occurs in the donor substrate.

The first concept shown in Figure~\ref{fig:lift_sketch}(a) is the simplest approach found in literature since the material ought to be transferred is also the layer where the laser energy is absorbed. Depending on the absorption of the material a complete or incomplete evaporation of the transfer material is possible~\cite{Wang2010},\cite{Wang2013c}. 

The second approach in Figure~\ref{fig:lift_sketch}(b) introduces a so-called sacrificial layer or dynamic release layer (DRL) between donor substrate and transfer layer~\cite{Nagel2008}. This layer is specifically designed to absorb the whole laser pulse energy, decompose, and build sufficient kinetic energy by vapor-pressure to remove the transfer layer from the donor without a large heat affection by the laser. One of the disadvantages is that with incomplete decomposition of the sacrificial layer debris of this material can be redeposited onto the receiver and/or transferred layer stack.

An approach to avoid this is shown in Figure~\ref{fig:lift_sketch}(c) which is sometimes called blister-assisted laser-induced forward transfer (BA-LIFT)~\cite{Kattamis2011}. Here, the sacrificial layer is only evaporated at the direct substrate/layer interface. The ultra fast expansion of the gas leads to a plastic deformation (blistering) of the sacrificial layer which accelerates the transfer layer kinetically to the receiver without any thermal load nor any debris redeposition~\cite{Miller2012}. 

The last example in Figure~\ref{fig:lift_sketch}(d1) and (d2) shows a similar topology as in the previous example with the difference that both donor and receiver substrate are in close contact with each other. Sometimes even certain pressure is applied to improve the transfer process. The transfer mechanism in this case can differ from the other examples since material shock-waves or thermal loading is sometimes required to drive the process~\cite{Ihlemann2014}.

There are many more approaches for LIFT or LDW. The concepts shown shall only give an idea about the approaches that could be feasible for the transfer of thin layer stacks in IBC SHJ solar cells. We believe that the topology shown in Figure~\ref{fig:lift_sketch}(c) is the most promising since the heat affection on the transfer layer is possibly the lowest in this case.

It is not clear whether there should be a gap between the donor and receiver or not (cf. Figure~\ref{fig:lift_sketch}(d1) and (d2)). Previous works have shown that the transfer process (or more specifically the flyer) is very sensitive to the gap distance and atmospheric conditions (i.e. sound velocity)~\cite{Fardel2010}. Furthermore, complex structures that can not be transfered with one laser pulse are most likely easier produced when there is no gap between the substrates~\cite{Ihlemann2014}. The transfer of a continuous interdigitated layer structure is not possible with a gap.

However, it is not clear if the commonly used interdigitated electrode pattern in IBC SHJ is the ideal structure in terms of efficiency. Studies indicate that a line- or point-like structure of the emitter and BSF regions could be very efficient~\cite{Chen2013a},\cite{Reinhold2011}. Such non-continuous, structures would be favorable for a LIFT process.

\begin{figure*}[htb]
	\centering
	\includegraphics[width=0.8\linewidth]{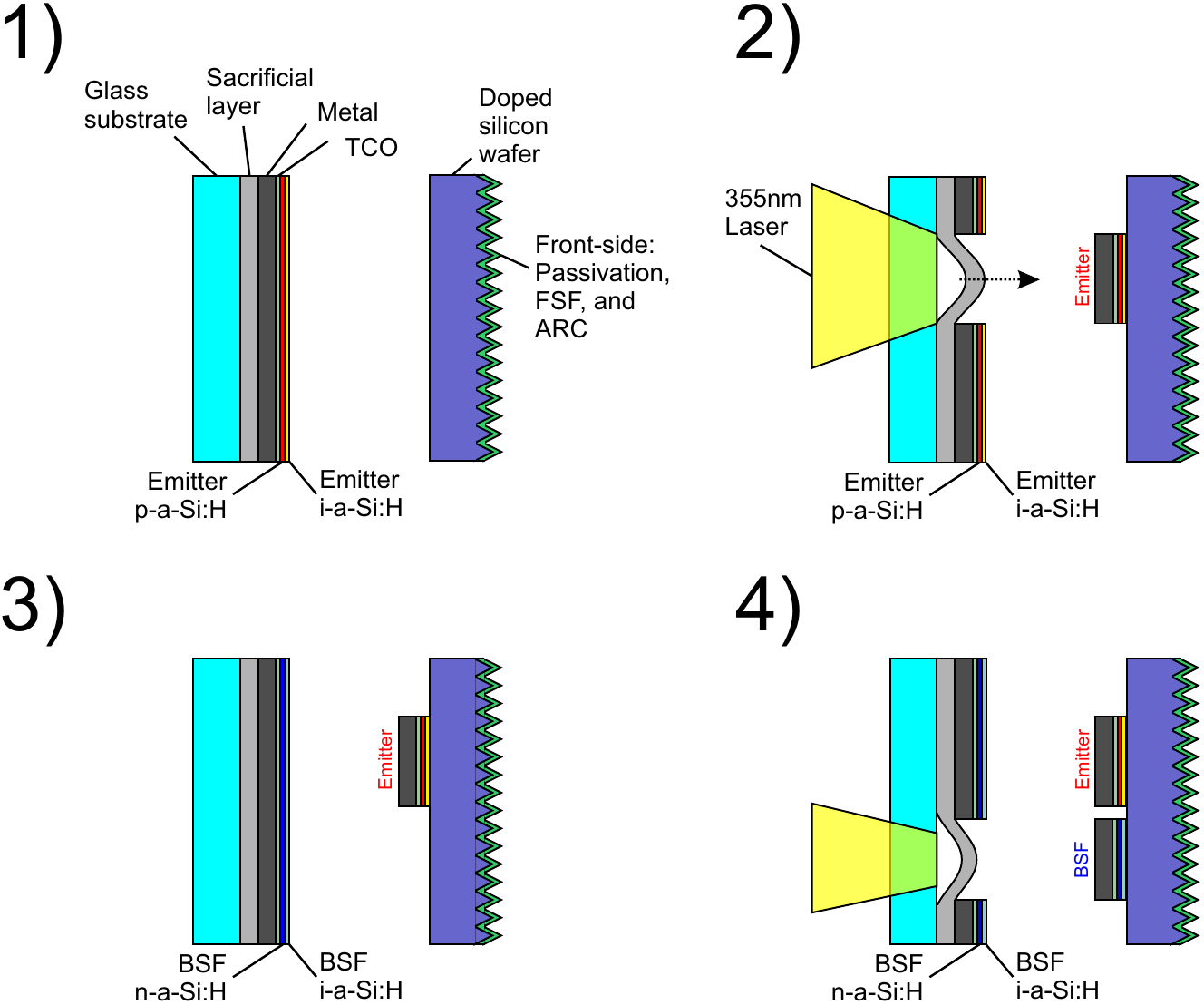}
	\caption{Blister-assisted LIFT for patterning of IBC SHJ solar cells. Process chain shows: 1) A donor substrate with the emitter region layer stack is placed on top of the receiver substrate, a crystalline silicon wafer with a processed front side. 2) A 355\,nm Q-switched DPSS nanosecond laser is used to illuminate the sacrificial layer which leads to partial evaporation and formation of high kinetic acceleration of the layer stack due to bulging. 3) A second donor substrate with the BSF layer stack is placed in front of the receiver. 4) Similar to step 2) a laser is used to transfer this layer stack appropriately. Alternatively, the intrinsic layer can be deposited beforehand onto the receiver if no differing deposition conditions for the emitter and BSF region respectively are needed.}
	\label{fig:lift_sketch_ibc}
\end{figure*}

Figure~\ref{fig:lift_sketch_ibc} shows the setup for the LIFT for patterning of IBC SHJ solar cells that is proposed in this work.

The sacrificial layer or DRL is deposited by spin-coating onto a glass substrate. Commonly used materials are polymers that are highly absorptive in the UV spectral range (i.e. 355\,nm). A polyimide of the type PI-2525 from HD Microsystems has proven to be suitable as such a layer~\cite{Miller2012}. Ideally, photo-induced decomposition of the polymer into volatile gases occurs under illumination by the laser. Sometimes also triazene and epoxy compounds are used for this task~\cite{Nagel2008}. In the proposed setup such properties are, ideally, not required since the blister remains intact.

\textbf{Preparation of the donor:} On top of the donor substrate with the sacrificial layer the function layer stack is deposited in reverse order. First, the metalization is deposited by RF magnetron sputtering of silver and aluminum-doped zinc-oxide. Afterwards, the doped hydrogenated amorphous silicon (a-Si:H) is deposited by plasma enhanced vapor deposition (PECVD)~\cite{Tanaka1992}. Optionally, also the intrinsic passivation layer is also deposited onto the donor with optimized properties for the respective region. Thus, two donor substrates are prepared. One for the emitter layer stack and one with the BSF layers. In every case all depositions are prepared homogeneously over the whole area of the substrate without any patterning. 

\textbf{Preparation of the receiver:} The receiver consists of a crystalline silicon wafer that is fully processed on the front-side and passivated by an intrinsic amorphous silicon layer (i-a-Si:H) on the back-side. Due to the low conductivity of i-a-Si:H a very thin layer of only approx. 5\,nm is used~\cite{Thibaut2011}. Thus, a sufficiently low resistance at the interface is achieved while at the same time a good passivation is ensured. As mentioned before, optionally the intrinsic layer is also transfered from the donor so no intrinsic layer is required on the receiver.

\textbf{The laser source:} In literature different lasers are used for the transfer depending on the processed materials. We believe that a Q-switched DPSS laser with a wavelength of 355\,nm and pulse durations in the nanosecond regime is suitable for the process. An arbitrarily laser beam intensity distribution is desirable to transfer the layers into an optimized pattern. Spatial light modulators (SLM) or deformable mirror devices (DMD) are capable of generating such shapes and have already been proven to be reliable for LIFT processes~\cite{Pique2014}. In any case, for a point-like emitter and BSF region transfer a top-hat intensity distribution could be sufficient.
It should be noted that due to the large areas of the transfered layers in the 100\,$\micro$m--1\,mm range a comparably high laser pulse energy ($\gg$1\,mJ) could be necessary. Furthermore, for a high parallelization of the process (i.e. by a SLM) even higher overall pulse energies are required.

\textbf{The transfer process:} After the donors and the receiver are prepared a suitable gap (or no gap) is chosen between both substrates. The atmospheric conditions are adjusted in a way that on the one hand, the flyer wont break or rupture by the recoil pressure of the air (for example lowered pressure). On the other hand, rupture on impact by high flyer velocities due to a low air resistance need to be avoided as well~\cite{Fardel2010}.

\section{Challenges}
As described in the transfer process, one of the main challenges is the determination of the possible process window, especially if there is a gap between the substrates. Although a heat affection of the transfer layers is unlikely, since a similar transfer process of whole layer stacks was reported before for the creation of OLED pixels~\cite{Constantinescu2014},\cite{Shaw-Stewart2011} and used in the OLED industry~\cite{Suh2003}, it is not clear how sensitive the very thin intrinsic/doped layers (10--20\,nm) are in terms of cracking and defect generation during and after the transfer. It is likely that organic semiconductors are less vulnerable against mechanical forces. Along with this arises the question about the impact of defects that are possibly generated at the edge of each transfered flyer. 

Another challenge is the adhesion of the layers on the receiver and especially the electrical and passivation properties of the formed heterojunction. It is possible that a no-gap transfer process is favorable with regards to this problem. However, proper transfer of a-Si:H onto crystalline silicon for doping was reported before which gives hope to believe that a good adhesion between amorphous and crystalline silicon layers is possible~\cite{Ferre2011}.

Lastly, with regards to technical difficulties, the resolution of the patterning designs is limited by the repeatability of the laser systems and accuracy of the LIFT onto the receiver. Modern laser material processing systems have a repeatability of the axes below 6\,$\micro$m. Fortunately, there is no alignment between donor substrates and the receiver required which could be a huge advantage over the commonly used patterning techniques (cf. Table~\ref{tab:comp}).  

The minimal flyer sizes are limited by the laser beam spot size. This could be important for the BSF area sizes. Literature shows that the gap distance between emitter and BSF should be as small as possible~\cite{Chen2013a}. A high geometrical fill-factor of the emitter is favorable~\cite{Haschke2013}. 

\section{Conclusion}
This work shall serve as a first publication that introduces the concept of laser-induced forward transfer (LIFT) for the patterning or printing of IBC SHJ cells. A small comparison of the possible LIFT topologies was given and a suitable setup for the present task was proposed. Possible challenges of transfer process and limitations specific for LIFT of amorphous silicon layers were discussed. Furthermore, challenges with regards to technical difficulties were mentioned as well. 

We believe that besides advantages with regards to process simplification and resolution improvement this method could enable a large degree of freedom in the choice of process parameters during the deposition of the solar cell. Since the whole back side emitter and BSF layer stacks are deposited individually fewer constraints on the deposition conditions (i.e. max. temperatures) are possible. When also the intrinsic passivation/tunneling layers are deposited by LIFT complete separately optimized processes for each region are possible. 

In the future experiments will be conducted that try to transfer doped layers onto passivated silicon wafers to investigate the adhesion as well as the electrical properties at the interface by conductivity measurements and lifetime mappings.

\begin{acknowledgements}
	The authors would like to thank U. Rau for fruitful discussions and support. 
\end{acknowledgements}

\Urlmuskip=0mu plus 1mu\relax
\bibliographystyle{spphys}       
\bibliography{all}   

\end{document}